In search of Einstein's brain

Steve C. Wang

Department of Mathematics and Statistics
Swarthmore College
Swarthmore, PA 19081
scwang@swarthmore.edu

In a cardboard box behind a beer cooler somewhere in Kansas lay Einstein's brain, apparently without any opinion. After its owner died in 1955, the brain was analyzed and interrogated, then left to its own resources, and since then has mostly been minding its own business. Every so often someone decides to awaken the brain from its meditation, seeking some small revelation about its late owner, and there arises a flurry of news about what will be found and learned from the endeavor. But the brain has not much to say, doesn't give up its secrets easily, and usually goes back to sleep none the worse for the wear.

Einstein and his brain have taken on a life of their own: like Elvis, Bigfoot, or kidnappings by aliens, they are a myth of our times. His brain is a symbol of a symbol, the very essence of genius personified. For Einstein holds a unique position among great minds: Newton and Darwin were scientists, Shakespeare a playwright, but Einstein was above all a genius. To be sure, he was also a scientist, but whereas Newton was a genius physicist,

Einstein was a physicist genius. Ask people who they associate with the word "genius," and they will invariably respond "Einstein."

Advertisements display his mythic status. The other day I noticed a rather tacky display featuring the stereotypical image of Einstein (flowing silver hair, warm but slightly goofy smile) and a caption reading, "Relatively speaking, [so-and-so] has the best prices on...." His likeness turns up in unlikely places: college dorm rooms commonly sport posters of Einstein, and I have even seen a T-shirt (an American icon itself) bearing his portrait. Recently, Einstein and his brain have even starred in that quintessential American genre, the road trip. In Michael Paterniti's book *Driving Mr. Albert*, the author recounts his saga with Dr. Thomas Harvey, the pathologist who performed the autopsy on Einstein, as the two journey across the country to meet Einstein's granddaughter, accompanied by the brain in a Tupperware container in their rented Buick Skylark. All this would be unthinkable for any mere scientist—seen any T-shirts of Niels Bohr lately? But for a cultural icon, for a mythic figure like Einstein, for a genius, it seems fitting.

As with any good myth, we harbor a generally cloudy understanding of Einstein and his work. We associate him with "genius" and "relativity" and various reports of his eccentricities; that is, with those traits that support our notion of Einstein as myth. But what of his science? How many know why he won his Nobel Prize? (It was not for relativity, but for the photoelectric effect in 1921.) We have but a fuzzy view of his work; we think of him merely pondering the universe and conjuring up some simple formula to describe it all. Writes Roland Barthes:

> Through the mythology of Einstein, the world blissfully regained the image of knowledge reduced to a formula.... Popular imagery expresses this: photographs of Einstein show him standing next to a blackboard covered with mathematical signs of obvious complexity; but cartoons of Einstein (the sign that he has become a legend) show him chalk still in hand, and having just written on an empty blackboard, as if without preparation, the magic formula of the world.

His was science of a special kind, science that dealt not just with minute details like electron levels of the atom or nucleotide sequences of DNA, but with our most basic notions of the universe and space and time—metaphysical physics, as it were, on the grandest level. His discoveries about relativity or the speed of light, although seemingly incomprehensible, were nonetheless ideas that anyone could contemplate. They possessed philosophical and even theological implications (God can go faster than the speed of light if He wants to, can't He?). It was science that made us think a little, and worry a little.

Consider the concept of the speed of light, the $c$ in $E = mc^2$, the fastest speed attainable: how must the world have felt when it came to know the meaning of an absolute, inviolable speed limit? This was the knowledge that humanity's infinite progress was not inevitable, and in a sense not possible: we could only go so fast. It meant that no matter how much our technology developed, even in an unimaginable future, we could never advance to the point at which we could go faster than the speed of light That point was not a function of technology; it was something that just couldn't be done. At the turn of the twentieth century, when Einstein proclaimed this cosmic speed limit, new forms of transportation were on the horizon: the automobile and the airplane, along with the familiar locomotive, held the promise of ever-speedier travel. An optimistic mind could easily picture machines that could go even faster—and why not infinitely fast? But when special relativity revealed $c$ hidden within the universe in such a fundamental way, how must that have affected a generation's dreams?

In many ways, Einstein's discoveries pushed us to question our deepest beliefs. So it was with many great scientific discoveries, from Copernicus's revelations about the solar system, to Darwin's insights on evolution by natural selection. Of course, that is precisely what makes these discoveries so great: forsaking all intuition, they supersede tenets previously held unquestionable. Who would have thought to imagine that the earth circled the sun, or that humans were distant cousins to worms? Who would have proposed a theory that portrayed space itself as curved through a fourth dimension, and accounted for the slowing of time? How could Einstein know these things—what went on inside his head?

Indeed, in the most literal sense, what did go on inside his head: how did that brain work? If we examine it, will we find it somehow magical? These questions must have been on the minds of many in April of 1955. On the eighteenth of that month, Einstein had died, and his son deemed that the vital organs were to be preserved for study; his father would have wanted it that way. As the brain was allowed to harden for a few days, the press speculated about what was to be found. Newsweek wondered about "The Great

Brain: Signs of Genius," and a New York Times headline noted "Key Clue Sought in Einstein Brain." According to the latter, previous studies had found no correlation between intelligence and the brain's mass or pattern of convolutions, but a 1926 study reported that compared to "half-wits," intelligent people had richer and more complex systems of blood vessels in the meninges. What then would the brain of Einstein, already considered the eminent genius, have in store for us?

That brain may have had a few surprises in it, but evidently it was holding on to them. Although reports of who was going to do the brain analyses and what they were looking for were plentiful in the weeks after Einstein's death, there are no subsequent reports of what anyone actually did or found. What happened? Did the analyses find nothing, and the media, reluctant to admit Einstein had an ordinary-looking brain, decide it was a non-story? Why don't any results about the brain examination appear in the Times or the major news periodicals, which earlier were all too happy to print optimistic stories about brain physiology?

Only decades after Einstein's death did findings begin to appear. For years, the brain had reposed in a cardboard cider box in Dr. Harvey's Kansas office, and little was heard from it. In 1985, however, a team of Berkeley neuroanatomists obtained several pieces of the brain. Their previous research had suggested that increased use of neuronal cells in rats (presumably a result of greater mental activity) would be reflected in an increased concentration of certain glial cells. Indeed, the group found a higher concentration of such cells in sections of Einstein's brain than in brains of a control group, although the difference was statistically significant in only one of the four sections they studied. Does this discovery hint at the origin of genius?

In 1999, forty-four years after Einstein's death, results appeared from the first study of gross anatomy of the brain. Researchers at McMaster University found that the parietal lobes, an area important in spatial and mathematical reasoning, were wider in Einstein's brain than in normal brains. Furthermore, the Sylvian fissure, a groove normally running through the parietal lobes, was partially absent—a unique feature that may have allowed the brain's neurons to more effectively connect with each other, the researchers conjectured. Overall, the brain turned out to be no heavier than normal (in fact, at 1230 grams, it weighed slightly less), but these morphological peculiarities were intriguing. "We don't have a clue, so anything that is suggested is interesting," commented a psychologist. "There must have been something about his brain that made him so brilliant."

If there is any meaning at all to be found in such analyses of Einstein's brain, it may tell us more about ourselves than about Einstein's brain. One must ask: why do we perform these analyses, and what do we really hope to find? Do we seek to add to our myth? We entertain the myth of Einstein the genius, sitting down at his desk, thinking for a while, and suddenly being struck with inspiration and discovering some simple formula describing the universe. Such a man's brain must be special in some way that we can measure. Maybe it's bigger or denser or has more blood vessels, but there must be something we can find. And when we find it, that will confirm our myth, won't it? It will all make sense then. It will give us a facile explanation for genius: it's all in the brain.

And then in the mediocrity of our lives, if we fail to rise occasionally to glorious heights, it will all be a result of sheer luck: we just didn't happen to be born with a brain like Einstein's. He was special—he was born that way. As for us, we do our best with what we have. We're just not Einsteins. It's not our fault. We're doing as best we can.

But we slice up Einstein's brain and try to find what it is that's special about it, where the difference lies, where the genius hides. And the brain, it twiddles its thumbs in its jar and goes on dreaming and minding its own business. If it is somehow special, it's not telling, and at night it sleeps peacefully, perhaps a little tired but content all the same.

*Acknowledgements*

I thank my college writing teachers: Dora Polachek, who introduced me to Roland Barthes and semiotics, and Stuart Davis, who nurtured my love of writing. This essay is the full version of a shorter piece that was published in Science as "In search of Einstein's genius" (289:5484, p. 1477, https://doi.org/10.1126/science.289.5484.1477).